# Designing Maintainable Hybrid Generative Systems: A Quantum-Inspired Approach to Automated Music Harmony Generation

***Josef Pavlíček***
*Czech Technical University in Prague /*
*Faculty of Information technology /*
*Department of Software Engineering*
*Prague/ Czech Republic* *josef.pavlicek@fit.cvut.cz*

**Abstract**

This paper presents the design and evaluation of a maintainable hybrid generative architecture for automated music harmony generation from melody. The proposed system combines quantum-inspired candidate exploration over overlapping melodic contexts with explicit rule-based optimization to balance generative flexibility and structural control. The architecture is evaluated using explicit and reproducible metrics covering structural coherence, functional agreement, harmonic similarity, and robustness. The results show that the proposed approach produces harmonizations that preserve tonal structure and cadential behavior while allowing multiple valid harmonic realizations. Furthermore, the optimization layer improves structural coherence, stability, and predictability without requiring a training corpus. The study demonstrates that transparent and controllable hybrid generative systems can be systematically designed and evaluated within the context of Information Systems Development.

**Keywords:** Hybrid Generative Systems, Quantum-Inspired AI, Music Harmony Generation, Rule-Based Optimization, Information Systems Development

## 1. Introduction

The problem of generating harmonic structures from melodic input has been studied across multiple domains, including music theory, cognitive science, artificial intelligence, and computational modeling. From early theoretical formulations of harmony based on mathematical relationships[1] and Pythagorean concepts of consonance [2], to modern computational approaches, harmonic organization has been understood as a structured system governed by both formal rules and perceptual constraints.

In traditional music theory, harmony is described in terms of functional relationships between chords, such as tonic, subdominant, and dominant roles [3], [4]. These functional relationships are closely tied to human perception of tonal stability and expectation[5], [6], [7]. From a cognitive perspective, harmonic processing can be interpreted as a decision-making process under constraints, where multiple valid alternatives may exist [8], [9].

In parallel, the field of artificial intelligence has developed methods for automatic harmonization and music generation. Early approaches include probabilistic models and rule-based systems [10], [11], while more recent work focuses on machine learning and deep neural networks [12], [13]. These approaches often aim to learn statistical patterns from large datasets and generate outputs that resemble existing musical styles.

However, purely data-driven approaches have limitations in terms of interpretability, controllability, and reproducibility. The generated outputs may be difficult to analyze, and the internal decision processes are often opaque. This creates challenges from an Information Systems Development (ISD) perspective, where transparency, maintainability, and explicit evaluation are essential[14], [15], [16].

Recent work has also explored the use of formal models and visualization techniques to better understand harmonic structure and reduce cognitive load [17], [18], [19]. These approaches highlight the importance of structured representations and explicit modeling of harmonic relationships.

In addition, alternative theoretical frameworks inspired by quantum theory have been proposed to model decision-making processes involving multiple coexisting alternatives [20], [21], [22]. While these models are not based on physical quantum computation, they provide useful conceptual tools for representing superposition, interference, and contextual decision dynamics.

Motivated by these perspectives, this paper proposes a hybrid generative system for automated harmonic generation. The system combines a generative module that explores multiple candidate harmonizations with a rule-based optimization layer that enforces structural and stylistic constraints.

Unlike approaches that aim to reproduce a single reference harmonization, the proposed system is designed to generate multiple valid harmonic alternatives that preserve functional structure and musical coherence. This reflects the nature of harmonic interpretation as a non-deterministic process with multiple acceptable solutions.

The main contribution of this work lies in the design and evaluation of a maintainable hybrid generative architecture within the context of Information Systems Development. The paper demonstrates that generative systems can be systematically evaluated using explicit and reproducible metrics, including structural properties, functional agreement, and harmonic similarity.

The remainder of the paper is organized as follows. Section 2 describes the experimental design and evaluation framework. Section 3 presents the results. Section 4 discusses the implications of the findings, and Section 5 concludes the paper.

## 2. Experimental Design

This section describes the experimental setup used to evaluate the proposed hybrid generative system. The goal of the experiment is to assess whether the rule-based optimization layer improves the structural quality, consistency, and reproducibility of generated harmonic outputs.

### 2.1. Research Questions

The evaluation is guided by the following research questions:

- **RQ1:** Does the rule-based optimization layer improve the structural quality of generated harmonic outputs?
- **RQ2:** Does the optimization layer reduce fragmentation and increase consistency across generated variants?
- **RQ3:** Can the quality of generated outputs be evaluated using explicit and reproducible metrics?

To answer the defined research questions, it is first necessary to describe the architecture of the evaluated system.

### 2.2. System Architecture

The proposed system follows a hybrid architecture that combines a generative exploration module with a rule-based optimization layer. From an Information Systems Development perspective, this design reflects principles of Design Science Research, where the system is developed as an artifact addressing the problem of controllable and interpretable harmonic generation. The overall system architecture is illustrated in Fig. 1.

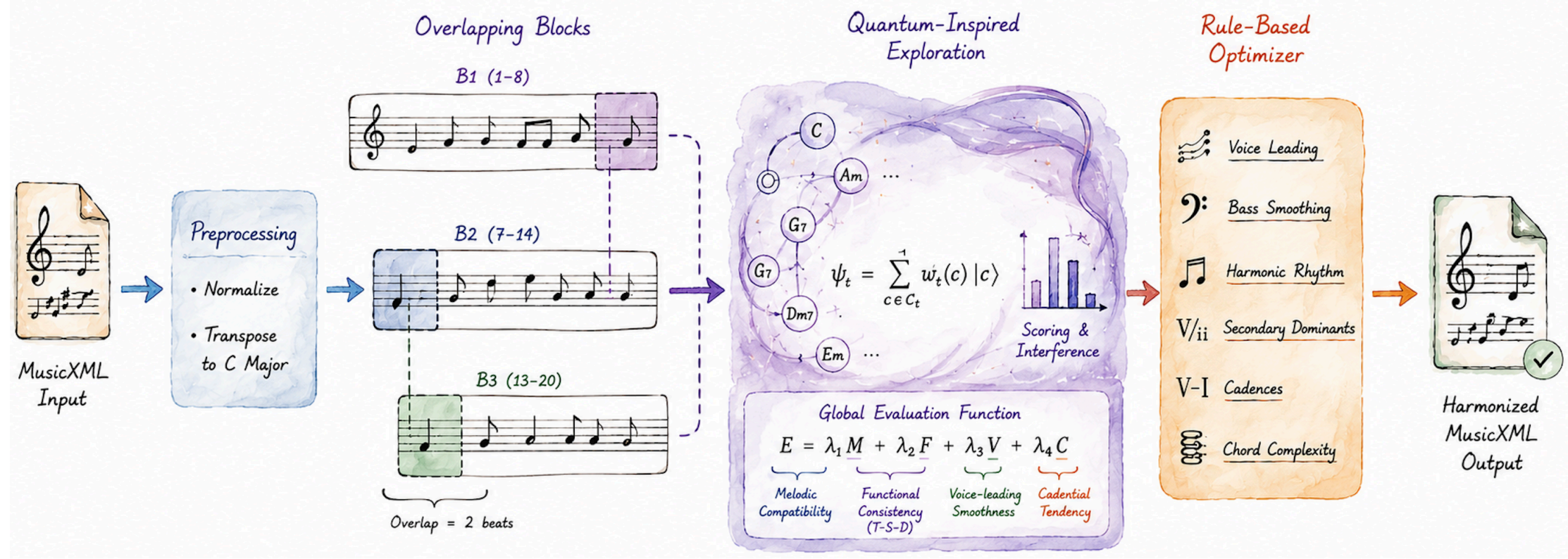

**Fig. 1.** A quantum-inspired approach to automated music harmony generation from melody through overlapping melodic context, candidate exploration, and rule-based optimization.

The architecture is modular and consists of two main components:

**(1) Generative Harmonizer** (Quantum-Inspired Module) - for each overlapping melodic block, the system generates multiple candidate chords for every beat. Rather than selecting a single chord deterministically, the generator maintains a weighted representation of all candidate chords at each time step. Formally, let $C_t$ denote the set of candidate chords for beat t. The candidate state is represented as:

$$\psi_t = \Sigma_{c\in C_t} w_t(c)|c\rangle,$$

where $w_t(c)$ denotes the current weight assigned to candidate chord c, and $|c\rangle$ is an abstract symbolic representation of that candidate. The notation is inspired by quantum mechanics and models a superposition of competing harmonic hypotheses rather than physical quantum states.

A complete harmonic realization is defined as a sequence of candidate chords:

$$H=(c_1,c_2,\dots,c_T),$$

which is evaluated using the global objective function:

$$E(H)=\Sigma_t (\lambda_1 M(c_t)+\lambda_2 F(c_t)+\lambda_3 V(c_t,c_{t-1})+\lambda_4 C(c_t)),$$

where $\lambda_1,\dots,\lambda_4$ are weighting coefficients controlling the relative importance of melodic compatibility, functional consistency, voice-leading smoothness, and cadential tendency.

Melodic compatibility is computed as:

$$M(c_t)=|N_t\cap P(c_t)|/|N_t|,$$

where $N_t$ denotes the melody notes active at beat t and $P(c_t)$ denotes the pitch classes contained in chord $c_t$.

Voice-leading smoothness is evaluated as:

$$V(c_t,c_{t-1})=1/(1+d(r_t,r_{t-1})),$$

where $d(r_t, r_{t-1})$ is the interval distance, measured in semitones, between consecutive chord roots.

Functional consistency $F(c_t)$ rewards valid tonic–subdominant–dominant progressions, while $C(c_t)$ assigns additional preference to cadential patterns near phrase endings.

After each iteration, the candidate weights $w_t(c)$ are updated according to the global score $E(H)$, reinforcing candidates that participate in high-scoring harmonic realizations while suppressing inconsistent alternatives. This iterative weight adaptation acts as an interference-like selection mechanism, gradually concentrating the candidate distribution toward structurally coherent harmonic solutions while efficiently exploring the large combinatorial search space induced by the overlapping melodic blocks.

**(2) Rule-Based Optimizer** (Post-processing Layer) - the optimization layer refines the generated harmonic sequence using explicit domain-specific rules. Unlike the generative module, which explores multiple possibilities, the optimizer enforces structural constraints to improve musical coherence.
The optimization includes:

- reduction of excessive chord changes within measures,
- smoothing of bass motion and voice leading,
- controlled insertion of harmonic functions (subdominant, dominant),
- introduction of secondary dominants and substitutions,
- adjustment of chord complexity (triads vs extended chords),
- cadence enforcement and stabilization of final harmonic resolution.

Importantly, the optimizer does not replace the generated harmonization entirely, but selectively modifies it when structural inconsistencies are detected.

### Architectural Implications

This separation between generative exploration and rule-based refinement represents a key architectural pattern. It improves system maintainability, as each component can be modified independently, and enhances transparency, since the decision process is partially interpretable through explicit rules.

### 2.3. System Variants

Two system configurations are evaluated:

- **Raw Generator:** The output of the generative harmonization module without post-processing.
- **Optimized Generator:** The output after applying the rule-based optimization layer.

To assess robustness, each configuration is executed multiple times with different random initializations.

### 2.4. Dataset

The evaluation dataset consists of eleven short monophonic melodies represented in MusicXML format. All melodies are normalized to a common tonal framework in C major to ensure comparability of harmonic behavior across examples. The melodic range is constrained to approximately two octaves, reflecting typical practical limitations of vocal or instrumental lines.

Although the dataset contains only eleven melodies, each composition represents a non-trivial harmonic search problem. The melodies contain approximately 16 measures on average, typically consisting of four quarter-notes per measure. Since harmonic candidates are evaluated at the beat level and include both triads and seventh chords, each melody generates a large combinatorial search space with exponentially growing numbers of possible harmonic realizations. For a typical melody of approximately 64 harmonic decision points and multiple chord candidates per position, the theoretical search space may easily exceed 10^50 alternative harmonic configurations. Consequently, the evaluation reflects not only variation across musical styles, but also the system's ability to navigate a substantial space of alternative harmonic solutions.

The dataset is intentionally constructed to cover a diverse set of harmonic contexts and stylistic patterns. It includes both well-known melodies and original compositions designed specifically for analytical purposes.

The dataset is composed of the following groups:

- Jazz standards (2 melodies): Autumn Leaves, Fly Me to the Moon
- Country-style melodies (2 melodies): Country Roads, Folsom Prison Blues
- Folk melodies (2 melodies): "Czech folk song" , It's a Long Way to Tipperary

Original compositions (5 melodies): Song I–V

The original compositions were designed to evaluate specific harmonic situations, including simple tonal structures (Song I), chord decomposition and major–minor transitions (Song II), cadential progression across scale degrees (Song III), and repetitive pop-style patterns (Songs IV–V).

**Each melody** is **associated with a reference harmonization**. The reference does not represent a single ground truth, but rather a target harmonic design that reflects a plausible and musically consistent realization. This allows evaluation of the generated outputs in terms of structural and functional similarity, rather than exact symbolic reproduction.

### 2.5. Experimental Procedure

For each melody in the dataset:

1. The system generates multiple harmonic sequences using different random seeds:
   a. *N* runs using the raw generator,
   b. *N* runs using the optimized generator.
2. For each generated sequence, a set of structural and comparative metrics is computed.
3. Results are aggregated across:
   a. multiple runs (to evaluate robustness),
   b. multiple melodies (to evaluate generality).

### 2.6. Evaluation Metrics

The evaluation is based on three groups of metrics, designed to capture structural properties, similarity to reference harmonizations, and system robustness.

**(A) Structural Output Metrics -** these metrics evaluate the internal structure of the generated harmonizations:

- **Chord Density:** average number of chord changes per measure.
- **Average Chord Duration:** mean duration of chord persistence.

- **Bass Jump:** average interval distance between consecutive bass notes.
- **Segment Length Standard Deviation:** variability of harmonic segment durations.

**(B) Reference-Based Metrics -** these metrics compare generated harmonizations with reference designs:

- **Exact Chord Match:** proportion of positions where the generated chord matches the reference exactly.
- **Functional Agreement:** similarity of harmonic functions (tonic, subdominant, dominant) between generated and reference harmonizations.
- **Harmonic Similarity:** pitch-class overlap between generated and reference chords.
- **Final Function Match:** agreement of the final harmonic function (cadential resolution).

**(C) Robustness Metrics -** to assess system stability, each configuration is evaluated across multiple runs under varying stochastic conditions. Robustness is measured using statistical properties of structural metrics:

- Mean and standard deviation of chord density,
- Mean and standard deviation of chord duration,
- Mean and standard deviation of bass jump.

To account for the fact that multiple valid harmonic realizations may exist for a given melody, the evaluation does not rely solely on exact chord matching. Instead, similarity measures based on harmonic function and pitch-class overlap are used to capture structural relationships between generated and reference harmonizations.

Harmonic Similarity is defined based on pitch-class overlap between generated and reference chords, allowing the comparison of harmonically related chords even when their symbolic representation differs (e.g., C vs Em or C vs Cmaj7).

Functional Agreement evaluates whether generated chords correspond to the same tonal function (tonic, subdominant, dominant) as the reference.

Exact Chord Match is retained as a strict baseline metric, but is not considered sufficient on its own.

### 2.7. Hypotheses

The experiment tests the following hypotheses:

- **H1:** The optimization layer increases effective harmonic density by reducing redundant chord repetition and introducing functionally meaningful harmonic transitions.
- **H2:** The optimization layer improves structural coherence by reducing bass movement discontinuities and decreasing variability of harmonic segments.
- **H3:** Optimized outputs preserve or improve functional agreement with reference harmonizations.
- **H4:** Optimized outputs maintain or improve harmonic similarity while improving structural coherence.

It should be noted that harmonic density is measured as the number of effective harmonic segments per measure after merging consecutive identical chords. Consequently, a higher density does not indicate a larger number of MusicXML harmony events, but rather a richer and less redundant harmonic structure.

### 2.8. Reproducibility

All experiments are conducted using fixed random seeds and a deterministic evaluation pipeline. The implementation, dataset, and evaluation scripts are made publicly available to ensure reproducibility.

## 3. Results

This section presents the quantitative evaluation of the proposed hybrid generative system. The results are organized into two main parts: (i) structural properties of the generated harmonizations, and (ii) comparison with reference harmonic designs.

### 3.1. Structural Output Metrics

Table 1 summarizes the structural characteristics of the generated harmonizations across the dataset.

**Table 1.** Structural Metrics (Raw vs Optimized)

| | Raw | Optimized |
|---|---|---|
| **Chord Density** | 1.5924 | 2.2673 |
| **Avg Chord Duration** | 2.6039 | 1.7840 |
| **Avg Bass Jump** | 3.5615 | 1.2165 |
| **Segment Length Std.** | 2.1593 | 1.3116 |

The results show a consistent effect of the optimization layer on the structure of the harmonic output. In all evaluated melodies, the optimized variant exhibits higher chord density compared to the raw generator. This indicates that the optimization process introduces additional harmonic articulation, resulting in a more finely structured progression. At the same time, the average chord duration decreases in the optimized outputs, reflecting a reduction in long, repetitive harmonic segments. Most importantly, the average bass jump is significantly reduced after optimization. This suggests that the optimization layer improves voice-leading smoothness and reduces abrupt harmonic transitions. Similarly, the standard deviation of segment lengths is reduced, indicating a more consistent and controlled harmonic structure.

These results confirm that the optimization layer does not merely simplify the output, but actively restructures it, producing harmonizations that are more balanced and musically coherent.

### 3.2. Reference-Based Evaluation

In addition to structural metrics, the generated harmonizations were evaluated with respect to reference harmonic designs. The results of the reference-based evaluation are summarized in Table 2.

**Table 2.** Reference-Based Evaluation

| | Raw | Optimized |
|---|---|---|
| **Exact Match** | 2.15% | 4.33% |
| **Functional Agreement** | 57.95% | 57.98% |
| **Harmonic Similarity** | 50.90% | 49.07% |
| **Final Function Match** | 90.91% | 90.91% |

The increase in chord density should not be interpreted as an increase in the number of explicit harmony events. Instead, it reflects a higher number of effective harmonic segments after redundant chord repetitions have been merged.

Exact chord matching remains low in both configurations (2.15% for raw and 4.33% for optimized outputs), confirming that the system does not aim to reproduce reference harmonizations directly.

Functional agreement remains stable at approximately 58% for both variants, indicating that the system consistently captures the underlying tonal structure of the reference harmonization.

Harmonic similarity, defined as pitch-class overlap between chords, reaches approximately 50% for both configurations. This suggests that even when chord symbols differ, the generated harmonies often share substantial tonal content with the reference.

Finally, cadence preservation remains high (approximately 91%), demonstrating that the system reliably maintains global harmonic structure.

Interestingly, the optimization stage does not significantly increase similarity to the reference. Instead, it preserves functional structure while restructuring harmonic details according to internal rule-based constraints.

### 3.3. Robustness Analysis

To assess the stability of the system, the structural metrics were analyzed across multiple runs with different initializations. The robustness metrics across repeated runs are summarized in Table 3.

**Table 3.** Robustness Metrics (mean ± standard deviation across runs)

| | Raw | Optimized |
|---|---|---|
| **Chord Density** | 1.56 ± 0.28 | 2.28 ± 0.27 |
| **Avg Chord Duration** | 2.63 ± 0.52 | 1.78 ± 0.33 |
| **Bass Jump** | 3.64 ± 0.36 | 1.23 ± 0.20 |

The results show that the optimized configuration consistently exhibits lower variance compared to the raw generator, indicating improved stability and predictability of the generated outputs.

The reported values represent mean and standard deviation of structural metrics. **Specifically, chord density is measured as the number of chord changes per measure, average chord duration is expressed in beats, and bass jump is measured in semitones between consecutive bass notes.** While the mean reflects the average structural property, the standard deviation captures how much these values vary across repeated runs.

This confirms that the optimization layer not only improves structural quality, but also reduces variability across runs, resulting in more consistent harmonic behavior.

### 3.4. Interpretation of Results

The combined results reveal an important characteristic of the proposed system. Rather than reproducing a single reference harmonization, the hybrid architecture generates alternative harmonic realizations that preserve tonal function and cadential behavior while allowing variation at the chord-symbol level.

The optimization layer improves structural coherence through smoother voice leading, controlled harmonic rhythm, and more balanced harmonic progression. Furthermore, the robustness analysis indicates reduced variability for several key structural metrics, particularly chord duration and bass movement, suggesting increased stability and predictability of the generated outputs.

Overall, the results support the interpretation of harmonic generation as a structured decision-making process in which multiple candidate solutions are explored and refined according to explicit musical constraints, producing harmonically plausible outputs without enforcing direct replication of a reference harmonization.

## 4. Discussion

### 4.1 Comparison with Existing Approaches

To better position the proposed system within the existing body of harmony generation research, Table 4 provides a conceptual comparison of representative approaches. The comparison focuses on explainability, controllability, and dependency on training data, which are particularly relevant from an Information Systems Development perspective.

**Table 4.** Comparison of Representative Harmony Generation Methods

| Method | Core Paradigm | Explainable | Rule Control | Training Corpus Required |
|---|---|---|---|---|
| **Paiement et al. (2006)** | Probabilistic Harmonization | Partial | Limited | yes |
| **Music Transformer** | Deep Learning | no | no | yes |
| **Traditional Rule-Based Systems** | Heuristic Rules | yes | full | no |
| **Proposed System** | Quantum-Inspired Hybrid | yes | full | no |

Unlike existing approaches, the proposed system combines quantum-inspired candidate exploration with explicit rule-based optimization, enabling controllable generation without requiring a training corpus. The term quantum-inspired refers to mathematical concepts such as superposition-like candidate representations and interference-based selection mechanisms. Although all experiments were performed on classical hardware, the formulation was designed with future quantum implementation in mind, where superposition and entanglement-like relationships between harmonic candidates may further improve optimization. Because the evaluation dataset was designed to cover representative harmonic situations rather than to form a random statistical sample, no formal significance tests were performed. Future evaluation on larger benchmark datasets will enable statistical hypothesis testing and broader generalization of the reported findings.

### 4.2 Discussion of Results

The results of the evaluation provide several important insights into the behavior of the proposed hybrid generative system.

**First**, the findings confirm that the system should not be interpreted as a mechanism for reproducing reference harmonizations. Exact chord matching remains low across both configurations, even after optimization. This indicates that the system does not aim to replicate specific harmonic sequences, but instead generates alternative solutions within a structured harmonic space.

This behavior is consistent with the system design. The generative module explores multiple candidate harmonizations using a quantum-inspired strategy, while the optimization layer applies rule-based refinement grounded in harmonic principles. As a result, the system prioritizes structural coherence over direct imitation.

**Second**, the relatively stable level of functional agreement (approximately 58% as shown in Table 2) suggests that the system captures essential aspects of tonal organization. Even when chord symbols differ, the generated harmonizations preserve underlying functional roles such as tonic, subdominant, and dominant relationships. This supports the interpretation of harmonic generation as a structured decision-making process rather than a symbolic matching task.

The reported functional agreement is based on a strict symbolic and functional classification of chords. The metric does not account for perceptual similarity or common harmonic substitutions (e.g., C major and A minor). Consequently, the reported values should be interpreted as conservative estimates of tonal similarity rather than measures of perceived musical equivalence.

**Third**, harmonic similarity values around 50% indicate that generated chords frequently share significant pitch-class content with the reference harmonizations. This is particularly important in cases where different chord symbols represent closely related harmonic structures (e.g., C vs Cmaj7 or C vs Em), demonstrating that the system produces musically meaningful alternatives rather than arbitrary substitutions.

Interestingly, the optimization layer does not systematically increase similarity to the reference. In some cases, similarity slightly decreases after optimization. However, this behavior reflects the role of the optimizer, which introduces additional harmonic detail such as inversions, chord extensions, and substitutions, while improving voice-leading and rhythmic placement. As a result, the output may diverge from the reference symbolically, while becoming more musically coherent.

This leads to an important conclusion: similarity to a reference harmonization is not a sufficient indicator of quality in generative harmonic systems. Multiple valid harmonic realizations may exist for a given melody, and enforcing strict similarity may reduce musical diversity. The proposed evaluation framework addresses this limitation by combining structural metrics with functional and pitch-based similarity measures.

From a broader perspective, the proposed system can be positioned between existing approaches. Deep learning models such as Transformer-based architectures [12], [13] achieve strong generative performance but typically operate as black-box systems with limited interpretability and controllability. In contrast, visualization-oriented systems such as midiVERTO [19] or chord wheel–based approaches [23], and recent quantum-inspired harmonic modeling work [24], focus on understanding harmonic structure rather than generating it. The proposed hybrid architecture bridges this gap by combining exploratory generation with explicit rule-based control.

From an Information Systems Development perspective, the key contribution lies in the system's modular and maintainable design. The separation between generative exploration and rule-based optimization enables transparency, controllability, and extensibility. Furthermore, the use of explicit and reproducible evaluation metrics allows systematic comparison of system configurations, moving beyond purely subjective musical assessment.

**Finally**, the behavior of the optimization layer can be interpreted as approximating aspects of human harmonic reasoning. Rather than selecting chords based solely on local matching, the system evaluates harmonic sequences in a broader structural context, taking into account functional progression, voice leading, and stylistic constraints. This results in harmonizations that resemble human-like decision processes, where multiple alternatives are considered and refined according to domain-specific rules.

**Overall**, the results demonstrate that the proposed hybrid generative system produces structurally coherent and functionally meaningful harmonic outputs, while maintaining flexibility, diversity, and improved stability across generated solutions.

## 5. Conclusion

This **paper presented** the design and **implementation of a hybrid generative system for automatic harmonic generation** based on a given melody, combining **a quantum-inspired candidate generation** module with a rule-based optimization layer. The evaluation demonstrates that the proposed architecture produces harmonizations that are structurally coherent and functionally meaningful, while not relying on direct reproduction of reference harmonizations. The system preserves key aspects of tonal organization, including functional relationships and cadential behavior, even when generating alternative chord realizations. The results further show that exact chord matching is not an adequate evaluation criterion for generative harmonic systems. Instead, a combination of structural metrics, functional agreement, and pitch-based harmonic similarity provides a more informative and reproducible assessment of output quality. This reflects the fact that multiple valid harmonic realizations may exist for a given melody.

**From an Information Systems Development perspective, the main contribution lies in the design of a maintainable and controllable hybrid architecture**. The separation between generative exploration and rule-based refinement enables transparency, interpretability, and systematic evaluation, addressing key limitations of purely data-driven approaches. To support reproducibility and further research, the implementation, dataset, and evaluation scripts are publicly available in the ISD2026_QuantumOne artifact of the repository: https://github.com/JosefPavlicek/quantum-inspired-music-research.git.

**Overall,** the proposed system demonstrates that it is possible to balance flexibility and control in generative processes, producing diverse yet structurally consistent outputs.

Future work will focus on extending the system with support for modulation, richer harmonic vocabularies, and style-specific constraints, as well as incorporating user-guided interaction into the generation process. The approach demonstrates that hybrid architectures can provide a practical balance between generative flexibility and system-level control.

## 6. Acknowledgement

The system architecture diagram (Figure 1) was generated using an AI-based image generation tool based on the author’s specification and design.